\documentclass[12pt,a4paper]{article}

\usepackage{amsfonts}
\usepackage{amsmath}
\usepackage{amssymb}

\let\ssection=\section
\renewcommand{\section}{\setcounter{equation}{0}\ssection}
\newcommand\eq[1]{(\ref{#1})}
\newcommand\eqs[2]{(\ref{#1}--\ref{#2})}

\newcommand\mathC{\mkern1mu\raise2.2pt\hbox{$\scriptscriptstyle|$}
        {\mkern-7mu\rm C}}           
\newcommand{\mathR}{\mathbb{R}}      

\newcommand\vs[1]{\vspace{#1pt}}

\newcommand\displayE[4]{
    \begin{center}
    \framebox{\begin{minipage}[t]{#1cm}\vs{#2}{\ #4}\vs{#3}
                \end{minipage}}
    \end{center}}

\newcommand\bra[1]{\langle #1|\,}
\newcommand\ket[1]{\,|#1\rangle}
\newcommand\ketbra[1]{\ket{#1}\bra{#1}}
\newcommand\A{\hat A}

\newcommand{\ga}{\gamma}

\renewcommand\l{\lambda}

\renewcommand\a{\alpha}
\renewcommand\b{\beta}
\newcommand\de{\delta}

\newcommand\De{\Delta}
\newcommand{\Ga}{\Gamma}
\renewcommand{\O}{\Omega}
\newcommand\Si{\Sigma}
\newcommand\Hi{{\cal H}}
\newcommand\R{{\cal R}}

\newcommand{\map}{\rightarrow}

\newcommand\ie{{i.e.},}
\newcommand{\op}{{\rm op}}

\newcommand\LeftDB{[\mkern-3mu[}
\newcommand\RightDB{]\mkern-3mu]}
\newcommand\Val[1]{\LeftDB\,#1\,\RightDB}

\newcommand\SAin[1]{\mbox{``}A\,\varepsilon\,#1\mbox{''}}
\newcommand\Ain[1]{A\,\varepsilon\,#1}

\newcommand\V[1]{{\cal V}(\Hi_{#1})}
\newcommand\Set{{\bf Sets}}                     
\newcommand\SetH[1]{\Set^{{\V{#1}}^{\rm op}}}   
\newcommand\SetC[1]{\Set^{{#1}^{\rm op}}}       

\newcommand\Sub[1]{{\rm Sub}(#1)}             
\newcommand\name[1]{\ulcorner #1\urcorner}

\newcommand\ps[1]{\underline{#1}}

\newcommand\w{\mathfrak{w}}

\newcommand{\Sig}{\ps{\Sigma}}            
\newcommand{\SR}{\ps{{\mathR}^{\succeq}}} 

\begin{document}

\title{Topos Methods in the Foundations of Physics\footnote{Based on joint work with Andreas D\"oring.}}
\author{Chris J. Isham\footnote{c.isham@imperial.ac.uk}}
\date{April, 2010}
\maketitle


\section{Introduction}\label{Sec:Intro}

Over forty years have passed since I first became interested in
the problem of quantum gravity. During that time there have been
many diversions and, perhaps, some advances. Certainly, the
naively-optimistic approaches have long been laid to rest, and the
schemes that  remain have achieved some degree of stability. The
original `canonical' programme evolved  into loop quantum gravity,
which has become one of the two  major approaches. The other, of
course, is string theory---a scheme whose roots lie in the old
Veneziano model of hadronic interactions, but whose true value
became apparent only after it had been re-conceived as a theory of
quantum gravity.

However, notwithstanding these hard-won developments, there are
certain issues in quantum gravity that transcend any of the
current schemes. These involve deep problems of both a
mathematical and a philosophical kind, and stem from a fundamental
paradigm clash between general relativity---the apotheosis of
classical physics---and quantum physics.

In general relativity, space-time `itself' is modelled by a
differentiable manifold $\cal M$: a set whose elements are
interpreted as `space-time points'. The curvature tensor of the
pseudo-Riemannian metric on $\cal M$ is then deemed to represent
the gravitational field. As a classical theory, the underlying
philosophical interpretation is realist: both the space-time and
its points truly `exist'\footnote{At least, that would be the view
of unreconstructed, space-time substantivalists. However, even
purely within the realm of classical physics this position has
often been challenged, particularly by those who place emphasis on
the relational features that are inherent in general relativity.},
as does the gravitational field.

On the other hand, standard quantum theory employs a background
space-time that is fixed \emph{ab initio} in regard to both its
differential structure and its metric/curvature. Furthermore, the
conventional interpretation is thoroughly instrumentalist in
nature, dealing as it does with counter-factual statements about
what would happen (or, to be more precise, the probability of what
would happen) \emph{if} a measurement is made of some physical
quantity.

In regard to quantum gravity, the immediate question is:
\displayE{9}{3}{3}{\!\!How can such a formalism be applied to
space and time themselves?} Specifically, what could it  mean to
`measure' properties of space and/or time if the very act of
measurement requires a spatio-temporal background within which it
is made? And how can we meaningfully talk about the `probability'
of the results of such measurements? A related question is what
meaning, if any, can be ascribed to quantum superpositions of
eigenstates of properties of space, time, or space-time. Over the years, this
issue has been much discussed by Roger Penrose   who concludes that
the existing quantum formalism simply cannot be applied to space
and time, and that some  new starting point is needed.

Another twist is provided by the subject of \emph{quantum
cosmology}  which aims at applying quantum theory to the entire
universe. There is no \emph{prima facie} link between this
aspiration and the subject of quantum gravity other than that we
have become accustomed to discussing cosmology using various
simple solutions to the classical equations of general relativity.
However, irrespective of the link with quantum gravity proper, it
is still thought provoking to consider in general terms what it
means to apply quantum theory to the `entire' universe. Of course,
this might simply be a stupid thing to do, but if one does attempt
it, then the problem of implementing instrumentalism becomes
manifest: for where is the external observer if the entire
universe is the system? In contexts like these one can see
clearly the attractions of finding a more realist interpretation
of quantum theory.

The complexity of such questions is significantly enhanced by the
lack of any experimental data that can be unequivocally identified
as pertaining to the subject of quantum gravity. In normal
theoretical physics there is the (unholy) trinity of (i)
real-world data; (ii) mathematical framework; and (iii)
conceptual/philosophical framework. These three factors are
closely interrelated in any real theory of the natural world:
indeed, this tripartite structure underpins all theoretical
physics.

However, in quantum gravity the first factor is largely missing,
and this raises some curious questions. In particular:
\begin{itemize}
\item Would we recognise the/a `correct' theory of quantum gravity even if it was handed to us on a plate?
Certainly, the Popperian notion of refutation is hard to apply
with such a sparsity of empirical data.

\item What makes any particular idea a
`good' one as far as the community is concerned? Relatedly, how is
it that one particular research programme becomes well established
whereas another falls by the wayside or, at best, gains only a
relatively small following?
\end{itemize}
The second question is not just whimsical, especially for our
younger colleagues, for it plays a key role in decisions about the
award of research grants, post-doc positions, promotion, and the like.

In practice, many of the past and present research programmes in
quantum gravity have been developed by analogy with other theories and
theoretical structures, particularly standard quantum field theory
with gauge groups. And as for why it is that certain ideas survive
and other do not,  the answer is partly that individual scientists
indulge in their own philosophical prejudices, and partly that they like using theoretical tools with which   they are already
familiar. This, after all, is the fastest route to writing new papers.

At this point it seems appropriate to mention the ubiquitous,
oft-maligned `Planck length'. This fundamental unit of length
comes from combining Planck's constant $\hbar$ (the `quantum' in
quantum gravity) with Newton's constant $G$ (the `gravity' in
quantum gravity) and the speed of light $c$ (which always lurks
around) in the form
\begin{equation}
    L_P:=\left(\frac{G\hbar}{c^3}\right)^{1/2}
\end{equation}
which has a value of around $10^{-35}$ meters; the corresponding
Planck time (defined as $T_P:=L_P/c$) has a value or around
$10^{-42}$ seconds.

The general assumption is that something `dramatic' happens to the
nature of space and time at these fundamental scales. Precisely
what that dramatic change might be has been the source of endless
speculation and conjecture. However, there is a fairly wide-spread
anticipation that in so far as spatio-temporal concepts have any
meaning at all in the `deep' quantum-gravity regime, the
appropriate mathematical model will not be based on standard,
continuum differential geometry. Indeed, it is not hard to
convince oneself that, from a physical perspective, the important
ingredient in a space-time model is not the `points' in that
space, but rather the `regions' in which physical entities can
reside. In the context of a topological space, such regions are
best modelled by open sets: the closed sets may be too
`thin'\footnote{ In a differentiable manifold, closed sets include
points and lines whereas physical  entities `take up room'.} to contain a
physical entity, and the only physically-meaningful closed sets
are those with a non-empty interior. These reflections lead
naturally to the subject of `pointless topology' and the theory of
\emph{locales}---a natural step along the road to topos theory.

However,   another frequent conjecture is that what we normally
call space and time (or space-time) will only `emerge' from the
correct quantum gravity formalism in some (classical?) limit. Thus
a fundamental theory of quantum gravity may (i) have no intrinsic
reference at all to spatio-temporal concepts; and (ii) be such
that some of the spatio-temporal concepts that emerge in various
limits are \emph{non-standard} and are modelled mathematically
with something other than topology and differential geometry.  All
this leads naturally to the main question that lies behind the
work reported in this chapter. Namely:
\displayE{12}{3}{3}{\!\!\!What is status of (or justification for)
using standard quantum theory when trying to construct a theory of
quantum gravity?}

\vspace{3pt} It is notable that the main current programmes in
quantum gravity \emph{do} all use essentially standard quantum
theory. However, around twelve  years ago I came to the conclusion
that the use of standard quantum theory was fundamentally
inconsistent, and  I stopped working in quantum gravity proper.
Instead, I  began studying what, to me, were the central problems
in quantum theory itself: a search that lead quickly to the use of
topos theory.

Of the various fundamental issues that arise, I will focus here on
just two of them. The first is the problem mentioned already:
viz, applying the standard instrumentalist interpretation of
quantum theory to space and time. The second is what I claim is a
category error in the use, \emph{a priori}, of the real and
complex numbers in the mathematical formulation of quantum theory
when applied in a quantum gravity context. However, to unpack this
problem it is first necessary to be more precise about the way
that the real (and complex) numbers arise in quantum theory. This
is the subject of the next  Section.

\section{The Problem of Using the Real Numbers in Quantum Gravity}
\label{Sec:RealNosinQG} The real numbers  arise in theories of
physics in three different (but related) ways: (i)  as the values
of physical quantities; (ii) as the values of probabilities; and
(iii) as a fundamental ingredient in models of space and time
(especially in those based on differential geometry). All three
are of direct concern vis-a-vis our worries about making
unjustified, \emph{a priori} assumptions in quantum theory. Let us
consider them in turn.

One reason for assuming physical quantities to be real-valued is
undoubtedly  grounded in the remark that, traditionally (\ie\ in
the pre-digital age), they are measured with rulers and pointers,
or they are defined operationally in terms of such measurements.
However, rulers and pointers are taken to be classical objects
that exist in the physical space of classical physics, and this
space is modelled using the reals. In this sense there is a direct
link between the space in which physical quantities take their
values (what we call the `quantity-value space') and the nature of
physical space or space-time \cite{Isham03}.

\displayE{12}{3}{3}{\!\!\!\!\! Thus assuming physical quantities
to be real-valued is problematic in any theory in which space, or
space-time, is not modelled by a smooth manifold. This is a
theoretical-physics analogue of the philosophers'  \emph{category
error}.}

Of course, real numbers also arise as the value space for
probabilities  via the relative-frequency interpretation of
probability. Thus, in principle, an experiment is to be repeated a
large number, $N$, times, and the probability associated with a
particular result is defined to be the ratio $N_i/N$, where $N_i$
is the number of experiments in which that particular result was obtained.
The rational numbers $N_i/N$ necessarily lie between $0$ and $1$,
and if the limit $N\map\infty$ is taken---as is appropriate for a
hypothetical `infinite ensemble'---real numbers in the closed
interval $[0,1]$ are obtained.

The relative-frequency interpretation of probability is natural in
instrumentalist theories of physics, but it is inapplicable in the
absence of any classical spatio-temporal background in which the
necessary sequence of measurements can be made (as, for example, is the
situation in quantum cosmology).

In the absence of   a relativity-frequency interpretation, the
concept of `probability' must be understood in a different way.
One possibility involves  the concept of `potentiality', or
`latency'. In this case there is no  compelling reason why the
probability-value space should be a subset of the real numbers.
The minimal requirement on this value-space is that it is an ordered
set, so that  one proposition can be said to be more or less
probable than another. However, there is no \emph{prima facie}
reason why this set should be \emph{totally} ordered.
\vspace{-7pt} \displayE{11}{3}{3}{\!\!In fact, one of our goals is
to dispense with probabilities altogether and to replace them with
`generalised' truth values for propositions.}

It follows from the above that, for us, a key problem is the way
in which the formalism of standard quantum theory is firmly
grounded in the concepts of Newtonian space and time (or the
relativistic extensions of these ideas) which are essentially
assumed \emph{a priori}. The big question is how this formalism
can be modified/generalised/replaced so as to give a framework for
physical theories that is
\begin{enumerate}
\item `realist' in some meaning of that word; and
\item not dependent \emph{a priori} on the real and/or complex numbers.
\end{enumerate}

For example, suppose we are told that there  is a background
space-time $\cal C$, but it is modelled on something other than
differential geometry; say, a causal set. Then what is the
quantum formalism that has the same relation to $\cal C$ as
standard quantum theory does to Newtonian space and time?

This question is very non-trivial as the familiar Hilbert-space
formalism is very rigid and does not lend itself to minor
`fiddling'. What is needed is something that is radically new and
yet which can still embody the basic principles\footnote{Of course, the question of what precisely are these `basic principles' is much debatable. I have a fond memory of being in the audience for a seminar by John Wheeler at a conference on quantum gravity in the early 1970s. John was getting well into the swing of his usual enthusiastic lecturing style and made some forceful remark about the importance of the quantum principle. At that point a hand was raised at the back of the lecture room, and a frail voice asked ``What \textit{is} the quantum principle?''. John Wheeler paused, looked thoughtfully at his interlocutor, who was Paul Dirac, and answered ``Well, to be honest, I don't know''. He paused again, and then said ``Do you?''. ``No'' replied Dirac. }of the quantum
formalism, or beyond. To proceed let us return to first principles
and consider what can be said in general about the general
structure of mathematical theories of a physical system.

\section{Theories of a Physical System}
\subsection{The Realism of Classical Physics}
\begin{quote}
\emph{``From the range of the basic questions of metaphysics we
shall here ask this \emph{one} question: What is a thing? The
question is quite old. What remains ever new about it is merely
that it must be asked again and again \cite{HeidThing}.''}
\end{quote}
\hfill{\bf Martin Heidegger} \vspace{10pt}

Asking such questions is not a good way for a modern young
philosopher to gain tenure. Plato was not ashamed to do so, and
neither was Kant, but at some point in the last century the
question became `Wittgensteined' and since then it is asked at
one's peril. Fortunately, theoretical physicists are not confined
in this way, and I will address the issue front on. Indeed, why
should a physicist be ashamed of this margaritiferous question? For is
it not what we all strive to answer  when we probe the physical
world?

Heidegger's own response makes interesting reading:
\begin{quote}
\emph{``A thing is always something that has such and such
properties, always something that is constituted in such and such
a way. This something is the bearer of the properties; the
something, as it were, that underlies the qualities.''}
\end{quote}
Let us see how modern physics approaches this fundamental question
about the beingness of Being.

In constructing a theory of any `normal'\footnote{That is, any
branch of physics other than quantum gravity!} branch of physics,
the key ingredients  are the mathematical representations of the
following:
\begin{enumerate}
\item Space, time, or space-time: the framework within which `things'
are made manifest to us.
\item `States' of the system: the mathematical entities that determine
`the way things are'.
\item Physical quantities pertaining to the system.
\item `Properties' of the system:  \ie\ propositions about
the values of physical quantities. A state assigns truth-values to
these propositions.
\end{enumerate}

In the case of standard quantum theory, the mathematical states are interpreted
in a non-realist sense as specifying only what would happen \emph{if}
certain measurements are made: so the phrase `the way things are'
has to be broadened to include this view. This
ontological concept must also apply to the `truth values' that are
\emph{intrinsically} probabilistic in nature. 

In addition, `the way
things are' is normally construed as referring to a specific
`moment' of time, but this could be generalised to be compatible
with whatever model of time/space-time is being employed. This
could even be extended to a `history theory', in which case `the
way things are' means the way things are for all moments of time,
or whatever is appropriate for the space-time model being
employed.

As theoretical physicists with a philosophical bent, a key issue is how such a
mathematical framework implies, or encompasses, various possible
philosophical positions. In particular, how does it interface with
the position of `realism'?

This issue is made crystal clear in the case of classical physics,
which is represented in the following way.
\begin{enumerate}
\item In Newtonian physics (which will suffice for illustrative purposes)
space is represented by the three-dimensional Euclidean space,
$\mathR^3$, and time by the one-dimensional space, $\mathR$.
\item To each system $S$ there is associated a set (actually a
symplectic differentiable manifold) $\cal S$ of states. At each moment of
time, $t\in\mathR$, the system has a unique state $s_t\in\cal S$.

\item  Any physical quantity, $A$, is represented by a function
$\breve{A}:{\cal S}\map\mathR$. The associated interpretation  is that if
$s\in\cal S$ is a state, then the value of $A$ in that state is
just the real number $\breve{A}(s)$. This is the precise sense in
which the philosophical position of (na\"ive) realism is encoded
in the framework of classical physics.

\item The basic propositions are of the form $\SAin\De$ which
asserts that the value of the physical quantity $A$ lies in the
(Borel) subset $\De$ of the real numbers. This proposition is
represented mathematically by the subset
$\breve{A}^{-1}(\De)\subseteq \cal S$---{\rm i.e.}, the collection
of all states, $s$,  in $\cal S$ such that $\breve{A}(s)\in\De$.
\end{enumerate}

This representation of propositions by subsets of the state space
$\cal S$ has a fundamental implication for the \emph{logical}
structure of classical physics: \displayE{11}{3}{3}{\!\!The
mathematical structure of set theory implies that, \emph{of
necessity}, the propositions in classical physics have the logical
structure of a \emph{Boolean algebra}.}

\vspace{10pt} We note that classical physics is the paradigmatic
implementation of Heidegger's view of a `thing' as the bearer of
properties. However, in quantum theory, the situation is very
different. There,  the existence of  any such realist
interpretation is foiled by the famous Kochen-Specker theorem
\cite{KS67}. This asserts that it is impossible to assign values
to all physical quantities at once if this assignment is to
satisfy the consistency condition that the value of a function of
a physical quantity is that function of the value. For example,
the value of `energy squared' is the square of the value of
energy.

\subsection{A Categorial Generalisation of the Representation of
Physical Quantities} The Kochen-Specker theorem implies that, from
Heidegger's world-view, there is \emph{no} `way things are'. To
cope with this, physicists have historically fallen back on the
instrumentalist interpretation of quantum theory with which we are
all so familiar. However, as I keep emphasising, in the context of
quantum gravity there are good reasons for wanting to achieve a
more realist view, and the central question is how this can be
done.

The problem is the great disparity
between the mathematical formalism of classical physics---which is naturally
realist---and the formalism of quantum physics---which is not. Let
us summarise these different structures as they apply to a
physical quantity $A$ in a system $S$ with associated propositions
$\SAin\De$ where $\De\subseteq\mathR$.
\paragraph{The classical theory of $S$:}
\begin{itemize}
\item The state space is a set $\cal S$.
\item $A$\ is represented by a function $\breve{A}:{\cal S}\map\mathR$.
\item The propositions $\SAin\De$ is represented by the subset
$\breve{A}^{-1}(\De)\subseteq\cal S$ of $\cal S$. The collection
of all such subsets forms a Boolean lattice.
\end{itemize}

\paragraph{The quantum theory of $S$:}
\begin{itemize}
\item The state space is a Hilbert space $\Hi$.
\item $A$ is represented by a self-adjoint operator, $\hat A$, on $\Hi$.
\item The proposition $\SAin\De$ is represented by the operator
$\hat E[\hat A\in\De]$ that projects onto the subset $\De\cap{\rm
sp}({\hat A})$ of the spectrum, ${\rm sp}(\hat A)$, of $\hat A$.
The collection of all projection operators on $\Hi$ forms a
\emph{non-distributive} lattice.
\end{itemize}

The `non-realism' of quantum theory is reflected in the fact that
propositions are represented by elements of a non-distributive
lattice, whereas in classical physics a distributive lattice
(Boolean algebra) is used.

So how can we find a formalism that goes beyond classical physics
and yet which retains some degree of realist interpretation? One
possibility  is to generalise the
axioms of classical physics  to a category, $\tau$, \emph{other}
than the category of sets. Such a representation of a physical
system would have the following ingredients:
\displayE{11}{4}{4}{{\bf The $\tau$-category theory of $S$:}
\begin{itemize}

\item There are two special objects, $\Si$, $\R$, in $\tau$ known respectively as
the \emph{state object} and \emph{quantity-value object}.

\item A physical quantity, $A$, is represented by an arrow
$\breve{A}:\Si\map\R$ in the category $\tau$.
\item Propositions about the physical world are represented by sub-objects of the state object $\Si$. In standard physics there must be some way of embedding into this structure propositions of the form $\SAin\De$ where 
$\Delta\subseteq\mathR$.
\end{itemize} }
But does such `categorification' work? In particular, is there
some category such that quantum theory can be rewritten in this
way?

\section{The Use of Topos Theory}
\subsection{The Nature of a Topos}
The simple answer to the question of categorification is `no', not
in general. The
sub-objects of an object in a general category do not have any logical structure, and I
regard possessing such a structure to be a \emph{sine qua non} for
the propositions in a physical theory. However, there is a special
type of category, known as a \emph{topos}, in which the
sub-objects of any object \emph{do} have a logical structure---a remark that underpins our entire research programme.

Broadly speaking, a topos, $\tau$, is a category that behaves in
certain critical respects just like the category, $\Set$, of sets.
In particular, these include the following:
\begin{enumerate}
\item There is an `initial' object, $0_\tau$, and a `terminal object',
$1_\tau$. These are the analogues of, respectively, the empty set,
$\emptyset$, and the singleton set $\{*\}$.

A \emph{global element} (or just \emph{element}) of an object $X$
is defined to be an arrow $x:1_\tau\map X$. This definition
reflects the fact that, in set theory, any element $x$ of a set
$X$ can be associated with a unique arrow\footnote{The singleton,
$\{*\}$ is not unique of course. But any two singletons are
isomorphic as sets and it does not matter which one we choose.}
$x:\{*\}\map X$ defined by\footnote{Note that, hopefully without
confusion, we are using the same letter `$x$' for the element in
$X$ and the associated function from $\{*\}$ to $X$.} ${x}(*):=x$.

The set of all global elements of an object $X$ is
denoted\footnote{In general, ${\rm Hom}_\tau(X,Y)$ denotes the
collection of all arrows in $\tau$ from the object $X$ to the
object $Y$.} $\Ga X$; {\rm i.e.}, $\Ga X:={\rm
Hom}_\tau(1_\tau,X)$.

\item One can form `products' and `co-products' of
objects\footnote{More generally, there are pull-backs and
push-outs.} These are the analogue of the cartesian product and
disjoint union in set theory.

\item In set theory, the collection, $B^A$, of functions $f:A\map B$ between
sets $A$ and $B$ is itself a \emph{set}; \ie\ it is an
object in the category of sets. In a topos, there is an analogous
operation known as `exponentiation'.  This associates to each pair
of objects $A,B$ in $\tau$ an object $B^A$ with the characteristic
property that
\begin{equation}
    {\rm Hom}_\tau(C,B^A)\simeq{\rm Hom}_\tau(C\times A,B)
    \label{Hom(C,BAA)}
\end{equation}
for all objects $C$. In set theory, the relevant statement is that
a parameterised family of functions $c\mapsto f_c:A\map B$, $c\in
C$, is equivalent to a single function $F:C\times A\map B$ defined
by $F(c,a):=f_c(a)$ for all $c\in C$, $a\in A$.

\item Each subset $A$ of a set $X$ is associated with a
unique `characteristic function' $\chi_A:X\map\{0,1\}$ defined by
$\chi_A(x):=1$ if $x\in A$ and $\chi_A(x)=0$ if $x\not\in A$.
Here, $0$ and $1$ can be viewed as standing for `false' and
`true' respectively, so that $\chi_A(x)= 1$ corresponds to the
mathematical proposition ``$x\in A$'' being true.

This operation is mirrored in any topos. More precisely, there is
a, so-called, `sub-object classifier', $\O_\tau$, that is the
analogue of the set $\{0,1\}$. Specifically,  to each sub-object $A$ of an
object $X$ there is associated a `characteristic arrow'
$\chi_A:X\map\O_\tau$; conversely, each arrow $\chi:X\map\O_\tau$ determines a unique
sub-object of $X$. Thus
\begin{equation}
    \Sub{X}\simeq{\rm Hom}_\tau(X,\O_\tau)
\end{equation}
where $\Sub{X}$ denotes the collection of all sub-objects of $X$.

\item The \emph{power set\/}, $PX$ of any set $X$ is
defined to be the set of all subsets of $X$. Each subset
$A\subseteq X$ determines, and is uniquely determined by, its
characteristic function $\chi_A:X\mapsto\{0,1\}$. Thus the set
$PX$ is in bijective correspondence with the function space
$\{0,1\}^X$. Analogously, in a general topos, $\tau$, we define
the \emph{power object} of an object $X$, to be $PX:=\O_\tau^X$.
It follows from \eq{Hom(C,BAA)} that
\begin{eqnarray}
\Ga(PX)&:=&{\rm Hom}_\tau(1_\tau,PX)={\rm
Hom}_\tau(1_\tau,\O_\tau^X)       \nonumber\\
      &\simeq&{\rm Hom}_\tau(1_\tau\times X,\O_\tau)\simeq
            {\rm Hom}_\tau(X,\O_\tau)\nonumber\\
      &\simeq&\Sub{X}
\end{eqnarray}
Since $\Sub{X}$ is always non-empty (because any object $X$ is
always a sub-object of itself) it follows that, for any $X$, the
object $PX$ is non-trivial.  As a matter of notation, if $A$ is a
sub-object of $X$ the corresponding arrow from $1_\tau$ to $PX$ is
called the `name' of $A$ and is denoted $\name{A}:1_\tau\map PX$.
\end{enumerate}

A key result for our purposes is that in any topos, the
collection, $\Sub{X}$, of sub-objects of any object $X$ forms a
\emph{Heyting algebra}, as does the set $\Ga\O_\tau:={\rm
Hom}_\tau(1_\tau,\O_\tau)$ of global elements of the sub-object
classifier $\O_\tau$. A Heyting algebra is a distributive lattice,
$\mathfrak{h}$, with top and bottom elements $1$ and $0$, and such
that if $\a,\b\in\mathfrak{h}$ there exists an element
$\a\Rightarrow\b$ in $\mathfrak{h}$ with the property
\begin{equation}
    \ga\preceq(\a\Rightarrow\b) \mbox{\ if and only if\ }
        \ga\land\a\preceq\b
\end{equation}
Then the `negation' of any $\a\in\mathfrak{h}$ is defined as
\begin{equation}
    \neg\a:=(\a\Rightarrow 0)
\end{equation}

A Heyting algebra has all the properties of a Boolean algebra
except  that the principle of \emph{excluded middle} may not hold.
That is, there may exist $\a\in\mathfrak{h}$ such that
$\a\lor\neg\a\prec 1$, \ie\ $\a\lor\neg\a\not= 1$.
Equivalently, there may be $\b\in\mathfrak{h}$ such that
$\b\prec\neg\neg\b$. Of course, in a Boolean algebra we have
$\b=\neg\neg\b$ for all $\b$. We shall return to this feature
shortly.

\subsection{The  Mathematics of `Neo-realism'}
Let us now consider the assignment of truth values to
propositions in mathematics. In set theory, let $K\subseteq X$ be a
subset of some set $X$, and let $x$ be an element of $X$. Then the
basic mathematical proposition ``$x\in K$'' is true if, and only
if, $x$ is an element of the subset $K$. At the risk of seeming
pedantic, the truth value, $\Val{x\in K}$, of this proposition can
be written as
\begin{equation}
\Val{x\in K}:=\left\{\begin{array}{ll}
            1 & \mbox{\ if\ $x$ belongs to $K$;} \\
            0 & \mbox{\ otherwise.}
         \end{array}
 \right.    \label{Def:nuxinK}
\end{equation}
In terms of the characteristic function $\chi_K:X\map\{0,1\}$, we
have
\begin{equation}
    \Val{x\in K}=\chi_K\circ{x}
    \label{nuxinKname}
\end{equation}
where $\chi_K\circ{x}:\{*\}\map\{0,1\}$.\footnote{For the
notation to be completely consistent, the left hand side of
\eq{Def:nuxinK} should really be written as $\Val{x\in K}(*)$.}

The reason for writing the truth value \eq{Def:nuxinK}  as
\eq{nuxinKname} is that this is the form that generalises in an
arbitrary topos. More precisely, let $K\in\Sub{X}$ with
characteristic arrow $\chi_K:X\map\O_{\tau}$, and let $x$ be a
global element of $X$ so that $x:1_\tau\map X$. Then the
`generalised' truth value of the mathematical proposition ``$x\in
K$'' is defined to be the arrow
\begin{equation}
    \Val{x\in K}:=\chi_K\circ{x}:1_\tau\map\O_\tau
    \label{nu(xinK)tau}
\end{equation}
Thus $\Val{x\in K}$ belongs to the Heyting algebra $\Ga\O_\tau$.
The adjective `generalised' refers to the fact that, in a generic
topos, the Heyting algebra contains elements other than just $0$
and $1$. In this sense, a proposition in a topos can be only
\emph{partially} true: a concept that seems ideal for application
to the fitful reality of a quantum system.

This brings us to our main contention, which is that it may be
profitable to consider constructing theories of physics in a topos
other than the familiar topos of sets
\cite{DI(1),DI(2),DI(3),DI(4),DI(Coecke)08,IB98,IB99,IB00,IB02}. The
propositions in such a theory admit a `neo'-realist interpretation
in the sense that there \emph{is} a `way things are', but this is specified by generalised truth
values  that may not be just true (1) or false (0).

Such a theory has the following ingredients:
\begin{enumerate}
\item A physical quantity $A$ is represented by a
$\tau$-arrow $\breve{A}:\Si\map\R$ from the state object $\Si$ to
the quantity-value object $\R$.
\item Propositions about $S$ are represented by elements of the
Heyting algebra, $\Sub{\Si}$, of sub-objects of $\Si$.  If $Q$ is
such a proposition, we denote by $\de(Q)$ the
associated sub-object of $\Si$; the `name' of $\de(Q)$ is the
arrow $\name{\de(Q)}:1_\tau\map P\Si$.

\item The topos analogue of a state is a `pseudo-state' which is a
particular type of sub-object of $\Si$. Given this pseudo-state,
each proposition can be assigned a truth value in the Heyting
algebra $\Ga\O_\tau$. Equivalently, we can use `truth objects':
see below.
\end{enumerate}
Conceptually, such a theory is neo-realist in the sense that the
propositions and their truth values belong to structures that
are `almost' Boolean: in fact they differ from Boolean algebras
only in so far as the principle of excluded middle may not apply.

Thus a theory expressed in this way `looks like' classical physics
except that classical physics always employs the topos $\Set$,
whereas other theories---including, we claim, quantum theory---use
a different topos. If the theory requires
a background space-time (or functional equivalent thereof) this
could be represented by another special object, $\cal M$, in the
topos. It would even be possible to mimic the actions of
differential calculus if the topos is such as to support
synthetic differential geometry\footnote{This approach
to calculus is based on the existence of
genuine infinitesimals in certain topoi.}.

The presence of intrinsic logical structures in a topos has another striking
implication. Namely, a topos can be used as a \emph{foundation}
for mathematics itself, just as set theory is used in the
foundations of `normal' (or `classical' mathematics). Thus
classical physics is modelled in the topos of sets, and thereby by
standard mathematics. But a theory of physics modelled in a topos,
$\tau$, other than $\Set$ is being represented in an alternative mathematical
universe!  The absence of excluded middle means that proofs by
contradiction cannot be used, but apart from that this, so-called,
`intuitionistic' logic can be handled in the same way as classical
logic.

A closely-related feature is that  each topos has an
`internal language' that is functionally similar to the formal
language on which set theory is based. It is this internal
language that is  used in formulating axioms for the mathematical
universe associated with the topos. The same language is also used
in constructing the neo-realist interpretation of the physical
theory.

\subsection{The Idea of a Pseudo-state}
In discussing the construction of truth values it is important to
distinguish clearly between truth values of \emph{mathematical}
propositions and truth values of \emph{physical} propositions. A
key step in constructing a physical theory is to translate the
latter into the former. For example, in classical physics, the
physical proposition $\SAin\De$ is represented by the subset
$\breve{A}^{-1}(\De)$\ of the state space, $\cal S$; {\rm i.e.},
\begin{equation}
\de(\Ain\De):=\breve{A}^{-1}(\De).
\end{equation}
Then, for any state $s\in\cal S$, the truth value of the physical
proposition $\SAin\De$ is defined to be the truth value of the
\emph{mathematical} proposition ``$s\in\de(\Ain\De)$'' (or,
equivalently, the truth value of ``$\breve{A}(s)\in\De$'').

Given the ideas discussed above, in a topos theory one might expect to
represent a physical state by a global element ${s}:1_\tau\map\Si$
of the state object $\Si$. The truth value of a proposition, $Q$,
represented by a sub-object $\de(Q)$ of $\Si$, would then be
defined as the global element
\begin{equation}
    \nu(Q;s):=\Val{s\in\de(Q)}=\chi_{\de(Q)}\circ{s}:1_\tau\map\O_\tau
    \label{nu(Q;s)}
\end{equation}
of $\O_\tau$. However, in the topos version of quantum theory (see
Section \ref{Sec:TopQT}) it transpires that $\Si$ has \emph{no global elements at
all}: in fact, this turns out to be precisely equivalent to
the Kochen-Specker theorem! This absence of global elements of the
state object could well be a generic feature of topos-formulated
physics, in which case we cannot use \eq{nu(Q;s)} and need to proceed in a
different way.

In set theory, there are two mathematical statements that are
equivalent to ``$s\in K$'''. These are:
\begin{equation}
\hspace{1pt}1.\hspace{3cm}    K\in T^s\hspace{7cm}                                                \label{OptionTruth}
\end{equation}
where the `truth object' $T^s$ is defined by\footnote{Note that
$s$ can be recovered from $T^s$ via
\begin{equation}
        \{s\}:=\bigcap_{K\in T^s}K
\end{equation}}
\begin{equation}
    T^s:=\{J\subseteq {\cal S}\mid s\in J\};
\end{equation}
and
\begin{equation}
\hspace{1pt}2.\hspace{3cm} \{s\}\subseteq K.\hspace{7cm}    
                                \label{OptionMacro}
\end{equation}
Thus, in classical physics, $\SAin\De$ is true in a state $s$ if,
and only, if (i) $\de(\Ain\De)\in T^s$; and (ii) $\{s\}\subseteq
\de(\Ain\De)$.

Let us consider in turn the topos analogue of these two options.
\paragraph{Option 1 (the truth-object option):} Note that
$T^s$ is a collection of subsets of $\cal S$: {\rm i.e.}, $T^s\in
\Sub{P{\cal S}}$. The interesting thing about
\eq{OptionTruth} is that it is of the form ``$x\in X$'' and
therefore has an immediate generalisation to any topos. More
precisely, in a general topos $\tau$ a truth object, $T$, would be
a sub-object of $P\Si$ (equivalently, a global element of
$P(P\Si)$)   with a characteristic arrow $\chi_T:P\Si\map\O_\tau$.
Then the physical proposition $Q$ has the topos truth value
\begin{equation}
    \nu(Q;T):=\Val{\de(Q)\in T}=\chi_T\circ\name{\de(Q)}
    :1_\tau\map\O_\tau
                \label{nu(Q;T)}
\end{equation}
The key remark is that although \eq{nu(Q;s)} is inapplicable if
there are no global elements of $\Si$, \eq{nu(Q;T)}
\emph{can} be used since global elements of a power object (like
$P(P\Si)$) always exist. Thus, in option 1, the analogue of a
classical state $s\in\cal S$ is played by the truth object $T\in
\Sub{P\Si}$.

\paragraph{Option 2 (the pseudo-state option):} We cannot use 
\eq{OptionMacro} in a literal
way in the topos theory since, if there are no global elements, $s$,
of $\Si$, trying to construct an analogue
of $\{s\}$ is meaningless. However, 
although there are no global elements of $\Si$, there may
nevertheless be certain sub-objects, $\w$, of $\Si$
(what we call `pseudo-states') that are, in some sense, as `close
as we can get' to the (non-existent) analogue of the singleton
subsets, $\{s\}$, of $\cal S$. We must then consider the
mathematical proposition ``$\w\subseteq \de(Q)$'' where $\w$ and
$\de(Q)$ are both sub-objects of $\Si$.

As we have seen in \eq{nu(xinK)tau}, the truth value of the
mathematical proposition ``$x\in K$'' is a global element of
$\O_\tau$: as such, it may have a value other than $1$
(`true') or $0$ (`false'). In other words, ``$x\in K$'' can be
only `partially true'. What is important for us is that an analogous situation arises if $J,K$ are
sub-objects of some object $X$. Namely,  a global
element, $\Val{J\subseteq K}$, of $\O_\tau$ can be assigned to the
proposition ``$J\subseteq K$'': \ie\ there is a precise
sense in which one sub-object of $X$ can be only `partially' a
sub-object of another. In this scenario, a physical
proposition $Q$ has the topos truth value
\begin{equation}
    \nu(Q;\w):=\Val{\w\subseteq\de(Q)}
\end{equation}
when the pseudo-state is $\w$. 
The general definition of $\Val{J\subseteq K}$ is not important
for our present purposes. In the quantum case, the explicit form  is discussed in the companion article by
my collaborator Andreas D\"oring \cite{Doe10}.

To summarise what has been said so far, the key ingredients in
formulating a theory of a physical system in a topos $\tau$ are the following:
\begin{enumerate}
\item There is a `state object', $\Si$, in $\tau$.
\item To each physical proposition, $Q$, there is associated
a sub-object, $\de(Q)$, of $\Si$.
\item The analogue of a classical state is given by either
(i) a truth object, $T$; or (ii) a pseudo-state $\w$. The topos
truth-value of the proposition $Q$ is then $\Val{\de(Q)\in T}$, or
$\Val{\w\subseteq \de(Q)}$, respectively.
\end{enumerate}

Note that no mention has been made here of the quantity-value
object $\R$. However, in practice, this also is  expected to play
a key role, not least in constructing the physical propositions. More precisely, we anticipate that each physical
quantity $A$ will be represented by an arrow $\breve{A}:\Si\map\R$,
and then a typical proposition will be of the form $\SAin\Xi$
where $\Xi$ is some sub-object of $\R$. 

If any `normal'
physics  is addressed in this way, physical quantities are
expected to be \emph{real}-valued, and the physical  propositions
are of the form $\SAin\De$ for some $\De\subseteq\mathR$.
In this case, it is necessary to decide what sub-object of $\R$ in
the topos corresponds to the external quantity
$\De\subseteq\mathR$.

There is subtlety here, however, since although (in any topos we
are likely to consider) there is a precise topos analogue of the
real numbers\footnote{Albeit defined using the analogue of Dedekind cuts,
not Cauchy sequences.}, it would be a mistake to assume that this is necessarily the quantity-value object: indeed, in our
topos version of quantum theory this is definitely \emph{not} the
case.

The question of relating $\De\subseteq\mathR$ to some sub-object
of the quantity-value object $\R$ in $\tau$ is just one aspect of
the more general issue of distinguishing quantities that are
\emph{external} to the topos, and those that are \emph{internal}.
Thus, for example, a sub-object $\Xi$ of $\R$ in $\tau$ is an
internal concept, whereas $\De\subseteq\mathR$ is external,
referring as it does to something, $\mathR$, that lies outside
$\tau$. Any reference to a background space, time or space-time
would also be external. Ultimately perhaps---or, at least,
certainly in the context of quantum cosmology---one would want to
have no external quantities at all. However, in any more `normal'
branch of physics it is natural that the propositions about the
system refer to the external (to the theory) world to which
theoretical physics is meant to apply. And, even in quantum
cosmology, the actual collection of what counts as physical
quantities is external to the formalism itself.

\section{The Topos of Quantum Theory}\label{Sec:TopQT}
\subsection{The Kochen-Specker theorem and contextuality}
To motivate our choice of topos for quantum theory let us return
again to the Kochen-Specker theorem which asserts the
impossibility of assigning values to all physical quantities
whilst, at the same time, preserving the functional relations
between them \cite{KS67}.

In a quantum theory, a physical quantity $A$ is represented by a
self-adjoint operator $\A$ on the Hilbert space, $\Hi$, of the
system. A `valuation' is defined to be a real-valued
function $\l$ on the set of all bounded, self-adjoint operators,
with the properties that: (i) the value $\l(\A)$ belongs to the
spectrum of $\A$; and (ii) the functional composition principle
(or FUNC for short) holds:
\begin{equation}
    \l(\hat B)=h(\l(\A))                  \label{funct-rule}
\end{equation}
for any pair of self-adjoint operators $\A$, $\hat B$ such that
$\hat B= h(\A)$ for some real-valued function $h$.

Several important results follow from this definition. For example, if $\A_1$ and $\A_2$ commute, it follows
from the spectral theorem that there exists an operator $\hat C$
and functions $h_1$ and $h_2$ such that $\A_1=h_1(\hat C)$ and
$\A_2=h_2(\hat C)$. It then follows from FUNC that
\begin{equation}
    \l(\A_1+\A_2)=\l(\A_1)+\l(\A_2)    \label{VA1+A2}
\end{equation}
and
\begin{equation}
    \l(\A_1\A_2)=\l(\A_1)\l(\A_2).\label{VAB}
\end{equation}

The Kochen-Specker theorem says that no valuations exist if
$\dim(\Hi)>2$. On the other hand, \eqs{VA1+A2}{VAB} show that, if it existed, a valuation restricted to 
a commutative sub-algebra of operators  would be just an element of the \emph{spectrum}\ of the algebra, and
of course such elements \emph{do} exist. Thus valuations
exist on any commutative sub-algebra\footnote{More precisely, since we
want to include projection operators, we assume that the
commutative algebras are von Neumann algebras. These algebras are
defined over the complex numbers, so that non self-adjoint
operators are included too.} of operators, but not on the (non-commutative) algebra,
${\cal B}(\Hi)$, of all bounded operators. We shall call such valuations `local'.

Within the instrumentalist interpretation of quantum theory, the
existence of local valuations is closely related to the
possibility of making `simultaneous' measurements on commutating
observables. However, the existence of local valuations also plays
a key role in the, so-called,   `modal' interpretations  in which
values are given to the physical quantities that belong to some
specific commuting set. The most famous such interpretation is
that of David Bohm where it is the configuration\footnote{If taken
literally, the word `configuration' implies that the state space
of the underlying classical system must be a cotangent bundle
$T^*Q$.} variables in the system that are regarded as always
`existing' (in the sense of possessing values).

The topos-implication of these remarks stems from the following
observations. First, let $V,W$ be a pair of commutative
sub-algebras with $V\subseteq W$.   Then any (local) valuation, $\l$,  on
$W$  restricts to give a valuation on the sub-algebra $V$. More
formally, if $\Sig_W$ denotes the set of all local valuations on
$W$, there is a `restriction map' $r_{WV}:\Sig_W\map\Sig_V$ in
which $r_{WV}(\l) :=\l|_V$ for all $\l\in\Sig_W$. It is clear that
if $U\subseteq V\subseteq W$ then
\begin{equation}
    r_{VU}(r_{WV}(\l))=r_{WU}(\l)\label{rVUrWV}
\end{equation}
for all $\l\in\Sig_W$---{\rm i.e.}, restricting from $W$ to $U$ is
the same as going from $W$ to $V$ and then from $V$ to $U$.

Note that, if one existed, a valuation, $\l$, on the non-commutative
algebra of all operators on $\Hi$ would provide an association of a
`local' valuation $\l_V:=\l|_{V}$ to each commutative algebra $V$ such
that, for all pairs $V,W$ with $V\subseteq W$ we have
\begin{equation}
    \l_W|_V=\l_V            \label{lWV=lV}
\end{equation}
The Kochen-Specker theorem asserts there are no such associations
$V\mapsto\l_V\in\Sig_V$ if $\dim{\Hi}>2$.

To explore this further consider the situation where we have three
commuting algebras $V,W_1,W_2$ with $V\subseteq W_1$ and
$V\subseteq W_2$ and  suppose 
$\l_1\in\Sig_{W_1}$ and $\l_2\in\Sig_{W_2}$ are local valuations. If there is  some commuting algebra $W$
so that (i) $W_1\subseteq W$ and $W_2\subseteq W$; and (ii) there exists
$\l\in\Sig_W$ such that $\l_1=\l|_{W_1}$ and $\l_2=\l|_{W_2}$ ,
then  \eq{rVUrWV} implies that
\begin{equation}
    \l_1|_{V}=\l_2|_{V}     \label{Match}
\end{equation}

However, suppose now the elements of $W_1$ and $W_2$\ do not all
commute with each other: \ie\ there is no $W$ such that
$W_1\subseteq W$ and $W_2\subseteq W$. Then although valuations
$\l_1\in\Sig_{W_1}$ and $\l_2\in\Sig_{W_2}$ certainly exist, there
is no longer any guarantee that they can be chosen to satisfy the matching
condition in \eq{Match}. Indeed, the Kochen-Specker theorem says
precisely that it is impossible to construct a collection of local
valuations, $\l_W$, for all commutative sub-algebras $W$ such that all
the matching conditions of the form \eq{Match} are satisfied.

We note that triples $V,W_1,W_2$ of this type arise when there are
non-commuting self-adjoint operators $\A_1,\A_2$ with a third operator $\hat B$ and functions
$f_1,f_2:\mathR\map\mathR$ such that $\hat B=f_1(\A_1)=f_2(\A_2)$.
Of course, $[\hat B,\A_1]=0$ and $[\hat B,\A_2]=0$ so that, in
physical parlance, $A_1$ and $B$ can be given `simultaneous
values' (or can be measured simultaneously) as can $A_2$ and $B$.
However, the implication of the discussion above is that the value
ascribed to $B$ (resp.\ the result of measuring $B$) depends on
whether it is considered together with $A_1$, or together with
$A_2$. In other words the value of the physical quantity $B$ is
\emph{contextual}. This is often considered one of the most
important implications of the Kochen-Specker theorem.

\subsection{The topos of presheaves $\SetH{}$}
To see how all this relates to topos theory let us rewrite the
above slightly. Thus, let $\V{}$ denote the collection of all
commutative sub-algebras of operators on the Hilbert space $\Hi$.
This is a partially-ordered set with respect to
sub-algebra inclusion. Hence it is also a category whose
objects are just the commutative sub-algebras of $\cal B(H)$; we shall call it the
`category of contexts'.

We view each commutative algebra as a context with which to
view the quantum system in  an essentially classical way in
the sense that the physical quantities in any such algebra can be
given consistent values, as in classical physics. Thus each context is a
`classical snapshot', or `world-view', or `window on reality'. In
any modal interpretation of quantum theory, only one context at a time is used\footnote{In the standard instrumentalist interpretation of quantum theory, a context is selected by choosing to measure a particular set of commuting observables.} but our intention is to use the collection of \emph{all}
contexts in one mega-structure that will capture the entire
quantum theory.

To do this, let us consider the association of the spectrum
$\Sig_V$ (the set of local valuations on $V$) to each commutative sub-algebra $V$.
As explained above, there are restriction maps
$r_{WV}:\Sig_W\map\Sig_V$ for all pairs $V,W$ with
$V\subseteq W$, and  these maps satisfy the 
conditions in \eq{rVUrWV}. In the language of category theory this means that  the
operation $V\mapsto\Sig_V$ defines the elements of a
\emph{contravariant functor}, $\Sig$, from the category $\V{}$ to the
category of sets; equivalently, it is a covariant functor from the
opposite category, $\V{}^\op$ to $\Set$. 

Now, one of the basic
results in topos theory is that for any category $\cal C$, the
collection of covariant functors $F:{\cal C}^\op\map\Set$ is a topos,
known as the `topos of presheaves' over $\cal C$, and is denoted
$\SetC{{\cal C}}$. In regard to quantum theory, our fundamental claim is that the
theory can be reformulated so as to look like classical physics, but
in the topos $\SetH{}$. The object $\Sig$ is known as the
`spectral presheaf' and plays a fundamental role in our theory. In purely mathematical terms this has
considerable interest as the foundation for a type of
non-commutative spectral theory; from a physical perspective, we
identify it as the state object in the topos.

The terminal object $1_{\SetH{}}$ in the topos $\SetH{}$ is the presheaf that associates
to each commutative algebra $V$ a singleton set $\{*\}_V$, and the
restriction maps are the obvious ones. It is then easy to see that
a global element $\l:1_{\SetH{}}\map\Sig$ of the spectral presheaf
is an association to each $V$ of a spectral element
$\l_V\in\Sig_V$ such that, for all pairs $V,W$ with
$V\subseteq W$ we have \eq{lWV=lV}. Thus we have the following
basic result: \displayE{11}{3}{3}{\!\!The Kochen-Specker theorem is
equivalent to the statement that the spectral presheaf, $\Sig$,
has no global elements.}

\vspace{5pt}
Of course, identifying the topos and the state object are only the
very first steps in constructing a topos formulation of quantum
theory. The next key step is to associate a sub-object of $\Sig$
with each physical proposition. In quantum theory, propositions
are represented by projection operators, and so what we seek is a
map
\begin{equation}
    \de:{\cal P}(\Hi)\map\Sub\Sig
\end{equation}
where ${\cal P}(\Hi)$ denotes the lattice of projection operators
on $\Hi$. Thus $\de$ is a map from a non-distributive quantum
logic to the (distributive) Heyting algebra, $\Sub\Sig$, in the
topos $\SetH{}$.

The precise definition of $\de(\hat P)$, $\hat P\in {\cal
P}(\Hi)$, is described in the companion article by  my
collaborator Andreas D\"oring \cite{Doe10}. Suffice it to say that, at each
context $V$, it involves approximating $\hat P$ with the `closest'
projector to $\hat P$ that lies in $V$: an operator that we call
`daseinisation' in honour of Heidegger's memorable existentialist perspective on
ontology.

The next step is to construct the quantity-value object, $\ps\R$.
We do this by applying the Gel'fand spectral transformation in
each context $V$. The most striking remark about the result is
that $\ps\R$ is \emph{not} the real-number object, $\ps\mathR$, in
the topos, although the latter is a sub-object of the former. One
result of this construction is that we are able to associate an
arrow $\breve{A}:\Sig\map\ps\R$ with each physical quantity $A$;
{\rm i.e.}, with each bounded, self-adjoint operator $\A$. This is
a type of non-commutative spectral theory.

The final, key, ingredient is to construct the truth objects, or
pseudo-states; in particular, we wish to do this for each vector
$\ket\psi$\ in the Hilbert space $\Hi$. Again, the details can be
found in the chapter by Andreas D\"oring, but suffice it to say that the
pseudo-state, $\ps\w^{\ket\psi}$ associated with $\ket\psi\in\Hi$
can written in the simple form
\begin{equation}
    \ps\w^{\ket\psi}=\de(\ketbra\psi)
\end{equation}
In other words, the pseudo-state corresponding to the unit vector $\ket\psi$ is just the daseinisation of the projection
operator onto $\ket\psi$.

\section{Conclusions}
We have revisited the oft-repeated statement about   fundamental incompatibilities between quantum theory and general relativity. In particular, we have argued that the conventional quantum formalism is inadequate to the task of quantum gravity both in regard to (i) the non-realist, instrumentalist interpretation; and (ii) the \emph{a priori} use of the real and complex numbers. 

Our suggestion is to employ a mathematical formalism that `looks like' classical physics (so as to gain some degree of `realism') but in a topos, $\tau$, other than the topos of sets. The ingredients in such a theory are:
\begin{enumerate}
\item A state-object, $\Si$, and a quantity-value object, $\R$.

\item A map $\de:{\cal P}\map\Sub\Si$ from the set of propositions, $\cal P$, to the Heyting algebra, $\Sub\Si$, of sub-objects of $\Si$.

\item A set of pseudo-states, $\w$, or truth objects, $T$. Then the truth value of the physical proposition $Q$ in the pseudo-state $\w$\ (resp.\ the truth object $T$) is $\Val{\w\subseteq \de(Q)}$ (resp.\ $\Val{\de(Q)\in T}$) which is a global element of the sub-object classifier, $\O_\tau$, in $\tau$. The set, $\Ga\O_\tau$ of all such global elements is also a Heyting algebra.
\end{enumerate}

In the case of quantum theory, in our published papers we have shown in great detail how this programme goes through with the topos being the category, $\SetH{}$, of presheaves over the category/poset $\V{}$ of commutative sub-algebras of the algebra of all bounded operators on the Hilbert space $\Hi$.

The next step in this programme is to experiment with `generalisations' of quantum theory in which the context category is not $\V{}$ but some category, $\cal C$, which has no fundamental link with a Hilbert space, and therefore with the real and/or complex numbers. In a scheme of this type the intrinsic contextuality of quantum theory is kept, but its domain of applicability could include spatio-temporal concepts that are radically different from those currently in use. 
\bigskip

{\bf Acknowledgements.} I am most grateful to Andreas D\"oring for the stimulating collaboration that we have enjoyed over the last four years.

\end{document}